\documentclass[11pt]{article}
\usepackage{amsmath}
\usepackage{amssymb}
\usepackage{theorem}
\usepackage{booktabs}
\usepackage{epsfig}
\usepackage{graphicx}
\usepackage{epstopdf}
\usepackage{centernot}
\usepackage[algoruled,linesnumbered,noend]{algorithm2e}

\DeclareMathOperator{\Pref}{Pref}
\DeclareMathOperator{\Suff}{Suff}
\DeclareMathOperator{\Card}{Card}

\DeclareMathOperator{\Fact}{Fact}

\newtheorem{proposition}{Proposition}[section]

\newtheorem{definition}{Definition}[section]
\newtheorem{theorem}{Theorem}[section]

\newtheorem{lemma}{Lemma}[section]

\newtheorem{example}{Example}[section]
\newtheorem{corollary}{Corollary}[section]

\newcommand{\PMCI}{\mathcal{PMC}}

\newcommand{\M}{\mathcal{M}}

\def\petitcarre{\vrule height4pt width 4pt depth0pt}
\def\enddim{\relax\ifmmode\eqno{\hbox{\petitcarre}}
\else
{\unskip\nobreak\hfil\penalty50
   \hskip2em\hbox{}\nobreak\hfil
   \petitcarre
   \parfillskip=0pt \finalhyphendemerits=0
  \par\medskip}\fi}

\def \begdim {\noindent {\sc Proof} : \par \noindent}

\DeclareMathOperator{\CFL}{CFL}
\DeclareMathOperator{\ICFL}{ICFL}

\DeclareMathOperator{\LCP}{LCP}
\DeclareMathOperator{\lcp}{lcp}

\numberwithin{equation}{section}

\title{\Large \bf Lyndon words versus inverse Lyndon words: queries on suffixes and bordered words}

\author{Paola Bonizzoni$^1$, Clelia De Felice$^2$, Rocco Zaccagnino$^2$, Rosalba Zizza$^2$ \\
$^1$Universit\`a degli Studi di Milano Bicocca\\
$^2$Universit\`a degli Studi di Salerno}

\begin{document}

\maketitle

\thispagestyle{plain}

\begin{abstract}
Lyndon words
have been largely investigated and showned to be a
useful tool to prove interesting combinatorial properties of words.
In this paper we state new properties of both Lyndon
and inverse Lyndon factorizations of a word $w$,
with the aim of exploring their use in some classical queries on $w$.

The main property we prove is related to a classical query on words. We prove
that there are relations between the length of the longest common extension (or longest common prefix) 
$\lcp(x,y)$ of two different suffixes $x,y$ of a word $w$ and the maximum length $\M$ of two consecutive factors of
the inverse Lyndon factorization of $w$. More precisely, $\M$ is
an upper bound on the length of $\lcp(x,y)$.  
This result is in some sense stronger than the compatibility property, proved by Mantaci, Restivo, Rosone
and Sciortino for the Lyndon factorization and here for the inverse Lyndon factorization. Roughly,
the compatibility property
allows us to extend the mutual order between local suffixes 
of (inverse) Lyndon factors to the suffixes of the whole word.

A main tool used in the proof of the above results is a property that we
state for factors $m_i$ with nonempty borders in an 
inverse Lyndon factorization: 
a nonempty border of $m_i$ cannot be a prefix of the next factor $m_{i+1}$.
The last property we prove shows that if
two words share a common overlap, then their Lyndon factorizations can be used to
capture the common overlap of the two words.

The above results open to the study of new applications of Lyndon words and
inverse Lyndon words in the field of string comparison.
\end{abstract}


\medskip
\noindent
$2010$ {\it Mathematics Subject Classification:} $68R15$, $68W32$.

\section{Introduction}

The Lyndon factorization of a word $w$ is a unique factorization of $w$ into a
sequence of Lyndon words in nonincreasing lexicographic ordering.
This factorization is one of the most known factorizations and 
it has been extensively studied
in different contexts, from formal languages to algorithmic stringology
and string compression. In particular the notion of a Lyndon word has been
shown to be useful in theoretical applications, such as the well known proof of
the {\em Runs Theorem}  \cite{Bannai17} as well in
string compression analysis. A connection between the Lyndon factorization and
the Lempel-Ziv (LZ) factorization has been given in \cite{Karkai17}, where it is shown that in general
the size of the LZ factorization is larger than the size of the
Lyndon factorization, and in any case the size of the Lyndon factorization cannot be
larger than a factor of 2 with respect to the size of LZ. This
result has been further extended in \cite{cpm19} to overlapping  LZ factorizations.
The Lyndon factorization has recently revealed to be a useful
tool also in investigating queries related to sorting suffixes of a word, with
strong potentialities for string comparison that have not been completely explored and understood
\cite{restivo-sorting-2014,MM13}.
Relations between Lyndon words and the Burrows-Wheeler Transform (BWT)
have been discovered first in \cite{CDP,MRRS} and more recently in \cite{varBWT}.
A main property of the Lyndon factorization is that it can be efficiently computed.
Linear-time algorithms for computing the Lyndon factorization
can be found in \cite{duval,FM} whereas an $\mathcal{O}(\lg{n})$-time parallel algorithm
has been proposed in \cite{apostolico-crochemore-1995,daykin}.

More recently Lyndon words found a renewed
theoretical interest and several variants of them have been
introduced and investigated with different motivations \cite{Nyldon2,DRR,DRR2}.
A related field studies the combinatorial
and algorithmic properties of {\it necklaces}, that are powers of Lyndon words,
and their prefixes or {\it prenecklaces} \cite{CRSSM}.
In \cite{inverseLyndon}, the notion of an inverse Lyndon word (a word which is
strictly greater than each of its proper suffixes)
has been introduced to define a new factorization,
called the {\em inverse Lyndon factorization}. 
A word which is not an inverse Lyndon word may
have different factorizations with inverse Lyndon words as factors
but each word $w$ admits a unique canonical inverse
Lyndon factorization, denoted $\ICFL(w)$.
This factorization has the property that a word
is factorized in a sequence of inverse Lyndon words,
in an increasing and {\em prefix-order-free} lexicographic ordering,
where prefix-order-free means that a factor cannot be a prefix of the consecutive one.
Moreover $\ICFL(w)$ can be still computed in linear time
and it is uniquely determined by $w$.

Differently from Lyndon words, inverse Lyndon words may be bordered.
As a main result in this paper, we show that if a factor $m_i$ in $\ICFL(w)$
has a nonempty border, then such a border cannot be inherited by the consecutive factor,
since it cannot be the prefix of the consecutive factor $m_{i+1}$.
In other words, the longest
common prefix between $m_i$ and $m_{i+1}$ is shorter than the border of $m_i$.
This result is proved by a further investigation on the connection between the Lyndon
factorization and the canonical inverse Lyndon factorization of a word, given in \cite{inverseLyndon}
through the {\em grouping} property. Indeed, given a word $w$
which is not an inverse Lyndon word, the factors in $\ICFL(w)$
are obtained by grouping together consecutive factors of the anti-Lyndon factorization
of $w$ that form a chain for the prefix order.

Thanks to the properties of $\ICFL(w)$,
the longest common extensions (or longest common prefix)
of two distinct factors in $\ICFL(w)$ appear
to have different properties than in the Lyndon factorization.  
In this framework, a natural question is whether and how the longest common extensions
of two factors of $w$ are related to the size of the factors in $\ICFL(w)$. 
We prove
that there are relations between the length of the longest common extension (or longest common prefix)
$\lcp(x,y)$ of two different factors $x,y$ of a word $w$ and the maximum length $\M$ of two consecutive factors of
the inverse Lyndon factorization of $w$. More precisely, $\M$ is
an upper bound on the length of $\lcp(x,y)$.  
This result is in some sense stronger than the compatibility property, proved in \cite{restivo-sorting,restivo-sorting-2014}
for the Lyndon factorization and here for the inverse Lyndon factorization.
Roughly, the compatibility property
allows us to extend the mutual order between local suffixes
of (inverse) Lyndon factors to the suffixes of the whole word.
Another natural question is the following.
\begin{quote}
{\em Given two words having a common overlap,
can we use their Lyndon factorizations to capture the similarity of these words? } 
\end{quote}
A partial positive answer to this question is provided here:
given a word $w$ and a factor $x$ of $w$, we prove that their Lyndon factorizations
share factors, except for the
first and last term of the Lyndon factorization of $x$.

The paper is organized as follows.
In Sections \ref{prel}, \ref{ILW}, \ref{icfl}, \ref{Group}, we gathered
the basic definitions
and known results we need.
Relations between the Lyndon factorizations of two words that
share a common overlap are proved in Section \ref{fact}.
Borders of inverse Lyndon words are discussed in Section \ref{Bord}.
The compatibility property for $\ICFL(w)$ is proved 
in Section \ref{SC}. Finally the upper
bound on the length of the longest common prefix of two factors 
of $w$ in terms of factors in $\ICFL(w)$ is stated in Section \ref{ss}.


\section{Preliminaries} \label{prel}

For the material in this section see
\cite{bpr,CK,Lo,lothaire,reu}.

\subsection{Words}

Let $\Sigma^{*}$ be the {\it free monoid}
generated by a finite alphabet $\Sigma$
and let $\Sigma^+=\Sigma^{*} \setminus 1$, where $1$ is
the empty word.
For a set $X$, $\Card(X)$ denotes the cardinality of $X$.
For a word $w \in \Sigma^*$, we denote by $|w|$ its {\it length}.
A word $x \in \Sigma^*$ is a {\it factor} of $w \in \Sigma^*$ if there are
$u_1,u_2 \in \Sigma^*$ such that $w=u_1xu_2$.
If $u_1 = 1$ (resp. $u_2 = 1$), then $x$ is a {\it prefix}
(resp. {\it suffix}) of $w$.
A factor (resp. prefix, suffix) $x$ of $w$
is {\it proper} if $x \not = w$.
Given a language $L \subseteq A^*$, we
denote by $\Pref(L)$ (resp. $\Suff(L)$, $\Fact(L)$) the set of all prefixes (resp.
suffixes, factors) of its elements.
Two words $x,y$ are {\it incomparable} for the prefix order, denoted as $x \Join y$,
if neither $x$ is a prefix of $y$ nor $y$ is a prefix of $x$.
Otherwise, $x,y$ are {\it comparable} for the prefix order.
We write $x \leq_{p} y$ if $x$ is a prefix of $y$
and  $x \geq_{p} y$ if $y$ is a prefix of $x$.
The notion of a pair of words comparable (or incomparable) for the suffix order
is defined symmetrically.

We recall that two words $x,y$ are called {\em conjugate} if there exist words
$u,v$ such that $x=uv, y=vu$.
The conjugacy relation is an equivalence relation. A conjugacy class
is a class of this equivalence relation.
The following is Proposition 1.3.4 in \cite{lotVecchio}.

\begin{proposition} \label{equazione}
Two words $x,y \in \Sigma^+$ are conjugate if and only if there exists
$r \in \Sigma^*$ such that
\begin{eqnarray} \label{EQ1}
xr & = & ry
\end{eqnarray}
More precisely, equality (\ref{EQ1}) holds if and only if there exist
$u,v \in \Sigma^*$ such that
\begin{eqnarray}
x & = & uv, \quad y = vu, \quad r \in u(vu)^*.
\end{eqnarray}
\end{proposition}

A {\it sesquipower} of a word $x$ is a word $w = x^np$ where
$p$ is a proper prefix of $x$ and $n \geq 1$.
A nonempty word $w$ is
\textit{unbordered} if
no proper nonempty prefix of $w$ is a suffix of $w$.
Otherwise, $w$ is {\it bordered}.
A nonempty word $w$ is \textit{primitive} if
$w = x^k$ implies $k = 1$. An unbordered word is primitive.

The following is a part of Proposition 1.3.2 in \cite{lotVecchio}.

\begin{proposition} \label{nocode}
Two words $u, v \in \Sigma^+$ commute if and only if
they are powers of the same word.
\end{proposition}

\subsection{Lexicographic order and Lyndon words}

\begin{definition} \label{lex-order}
Let $(\Sigma, <)$ be a totally ordered alphabet.
The {\it lexicographic} (or {\it alphabetic order})
$\prec$ on $(\Sigma^*, <)$ is defined by setting $x \prec y$ if
\begin{itemize}
\item $x$ is a proper prefix of $y$,
or
\item $x = ras$, $y =rbt$, $a < b$, for $a,b \in \Sigma$ and $r,s,t \in \Sigma^*$.
\end{itemize}
\end{definition}

In the next part of the paper we will implicitly refer to
totally ordered alphabets.
For two nonempty words $x,y$, we write $x \ll y$ if
$x \prec y$ and $x$ is not a proper prefix of $y$
\cite{Bannai15}. We also write $y \succ x$ if $x \prec y$.
Basic properties of the lexicographic order are recalled below.

\begin{lemma} \label{proplexord}
For $x,y \in \Sigma^*$, the following properties hold.
\begin{itemize}
\item[(1)]
$x \prec y$ if and only if $zx \prec zy$,
for every word $z$.
\item[(2)]
If $x \ll y$, then $xu \ll yv$
for all words $u,v$.
\item[(3)]
If $x \prec y \prec xz$ for a word $z$,
then $y = xy'$ for some word $y'$ such that $y' \prec z$.
\end{itemize}
\end{lemma}

\begin{lemma} \label{proplexordbis}
Let $x,y \in \Sigma^*$.
If $x \ll y$, then $y \not \prec x$.
\end{lemma}
\begdim
Suppose, contrary to our claim, that
there would be $x,y \in \Sigma^*$ such that
$y \prec x \ll y$.
By definition there are $a,b \in \Sigma$ and $r,s,t \in \Sigma^*$ such that
$x = ras$, $y =rbt$.
Thus $y$ cannot be a prefix of $x$, hence there are $a',b' \in \Sigma$ and $r',s',t' \in \Sigma^*$
such that $y = r'a's'$, $x =r'b't'$, $a' < b'$.
By $x = ras = r'b't'$ we have that the words $ra$, $r'b'$ are comparable for the prefix order.
If $r'b'$ would be a prefix of $r$, then $r'b'$ were a prefix of $y = r'a's'$, which
is impossible. Analogously, if $ra$ would be a prefix of $r'$, then $ra$ were a prefix of $y = rbt$,
once again a contradiction.
Hence $ra = r'b'$, which implies $r = r'$, $a = b'$, therefore $a' = b > a = b' > a'$,
a contradiction.
\enddim

\begin{definition}\label{Lyndon-word}
A Lyndon word $w \in \Sigma^+$ is a word which is primitive and the smallest one
in its conjugacy class for the lexicographic order.
\end{definition}

\begin{example}
{\rm Let $\Sigma = \{a,b\}$ with $a < b$.
The words $a$, $b$, $aaab$, $abbb$, $aabab$ and $aababaabb$
are Lyndon words. On the contrary, $abab$, $aba$ and
$abaab$ are not Lyndon words. Indeed, $abab$ is a non-primitive word,
$aab \prec aba$
and $aabab \prec abaab$.}
\end{example}

Lyndon words are also called {\it prime words}
and their prefixes are also called {\it preprime words} in \cite{Knuth}.
Some properties of Lyndon words are recalled below.

\begin{proposition} \label{P1}
Each Lyndon word $w$ is unbordered.
\end{proposition}

\begin{proposition} \label{P2}
A word $w \in \Sigma^+$ is a Lyndon word if and only if
$w \prec s$, for each nonempty proper suffix $s$ of $w$.
\end{proposition}

The following is Proposition 5.1.3 in \cite{lotVecchio} and gives a second characterization
of Lyndon words.

\begin{proposition} \label{propP3}
A word $w \in \Sigma^+$ is a Lyndon word if and only if $w \in \Sigma$ or
$w = \ell m$ with $\ell, m$ Lyndon words, $\ell \prec m$.
\end{proposition}

Finally, in \cite{DRR} the authors credited to folklore the following third characterization of
Lyndon words:
$w \in \Sigma^+$ is a Lyndon word if and only if for each nontrivial
factorization $w = uv$ one has $u \prec v$.

A class of conjugacy is also called a {\it necklace} and often identified
with the minimal word for the lexicographic
order in it. We will adopt this terminology. Then
a word is a necklace if and only if it is a power of a Lyndon word.
A {\it prenecklace} is a prefix of a necklace. Then clearly
any nonempty prenecklace $w$ has the form $w = (uv)^ku$, where $uv$ is a Lyndon word,
$u \in \Sigma^*$, $v \in \Sigma^+$, $k \geq 1$, that is, $w$ is a sesquipower of a Lyndon word $uv$.
The following result has been proved in \cite{duval}.

\begin{proposition} \label{preprime}
A word is a nonempty preprime word if and only if it is a sesquipower of a Lyndon word distinct of
the maximal letter.
\end{proposition}

The proof of Proposition \ref{preprime} uses the following result which
characterizes, for a given nonempty prenecklace $w$ and a letter $b$, whether $wb$ is still
a prenecklace or not and, in the first case, whether $wb$ is a Lyndon word or not \cite[Lemma 1.6] {duval}.

\begin{theorem} \label{FundamentalPreneck}
Let $w = (uav')^ku$ be a nonempty prenecklace, where $uav'$ is a Lyndon word,
$u, v' \in \Sigma^*$, $a \in \Sigma$, $k \geq 1$. For any $b \in \Sigma$, the
word $wb$ is a prenecklace if and only if $b \geq a$. Moreover $wb \in L$ if and only if
$b > a$.
\end{theorem}

A direct consequence of Theorem \ref{FundamentalPreneck} is reported below
(see \cite[Theorem 2.1] {CRSSM} which states both Theorem \ref{FundamentalPreneck}
and Corollary \ref{FundamentalPreneckCor}).

\begin{corollary} \label{FundamentalPreneckCor}
Let $w = (uav')^ku$ be a nonempty prenecklace, where $uav'$ is a Lyndon word,
$u, v' \in \Sigma^*$, $a \in \Sigma$, $k \geq 1$. Let $b \in \Sigma$ with $b \geq a$
and let $y$ be the longest prefix
of $wb$ which is a Lyndon word. Then
$$y = \begin{cases} uav' & \mbox{ if }  b = a
\\
wb & \mbox{ if } b > a \end{cases}$$
\end{corollary}


\subsection{The Lyndon factorization} \label{LyFa}

A family $(X_i)_{i \in I}$ of subsets of $\Sigma^+$, indexed by a totally ordered set $I$, is
a {\it factorization of the free monoid} $\Sigma^*$ if each word $w \in \Sigma^*$
has a unique factorization
$w = x_1 \cdots x_n$, with $n \geq 0$, $x_i \in X_{j_i}$ and $j_1 \geq j_2 \geq \ldots \geq j_n$ \cite{bpr}.
A factorization $(X_i)_{i \in I}$ is called {\it complete} if each $X_i$ is reduced to a singleton
$x_i$ \cite{bpr}.
In the following $L = L_{(\Sigma^*, <)}$ will be the set of Lyndon words, totally
ordered by the relation $\prec$ on $(\Sigma^*, <)$.
The following theorem
shows that the family $(\ell)_{\ell \in L}$ is a complete factorization of $\Sigma^*$.

\begin{theorem} \label{Lyndon-factorization}
Any word $w \in \Sigma^+$ can be written in a unique way as
a nonincreasing product $w=\ell_1 \ell_2 \cdots \ell_h$ of Lyndon words, i.e., in the form
\begin{eqnarray} \label{LF}
w & = & \ell_1 \ell_2 \cdots \ell_h, \mbox{ with } \ell_j \in L \mbox{ and } \ell_1 \succeq \ell_2 \succeq \ldots \succeq \ell_h
\end{eqnarray}
\end{theorem}

The sequence $\CFL(w) = (\ell_1, \ldots, \ell_h)$ in Eq. (\ref{LF}) is called the
\textit{Lyndon decomposition} (or \textit{Lyndon factorization}) of $w$.
It is denoted by $\CFL(w)$ because Theorem \ref{Lyndon-factorization}
is usually credited to Chen, Fox and Lyndon
\cite{Lyndon}.
Uniqueness of the above factorization is a consequence of the following result, proved
in \cite{duval}.

\begin{lemma} \label{duval-prop}
Let $w \in \Sigma^+$ and let $\CFL(w) = (\ell_1, \ldots, \ell_h)$.
Then the following properties hold:
\begin{itemize}
\item[(i)]
$\ell_h$ is the nonempty suffix of $w$ which is the smallest with respect to the lexicographic order.
\item[(ii)]
$\ell_h$ is the longest
suffix of $w$ which is a Lyndon word.
\item[(iii)]
$\ell_1$ is the longest
prefix of $w$ which is a Lyndon word.
\end{itemize}
\end{lemma}

A direct consequence is stated below and it is necessary for our aims.

\begin{corollary} \label{corP4}
Let $w \in \Sigma^+$, let $\ell_1$ be its longest
prefix which is a Lyndon word and let $w'$ be
such that $w= \ell_1 w'$. If $w' \not = 1$, then $\CFL(w) = (\ell_1, \CFL(w'))$.
\end{corollary}

As a consequence of Theorem \ref{Lyndon-factorization}, for any word $w$ there is a factorization
$$w = \ell_1^{n_1} \cdots \ell_r^{n_r}$$
where $r  > 0$, $n_1, \ldots , n_r \geq 1$, and $\ell_1 \succ \ldots \succ \ell_r$
are Lyndon words, also named {\it Lyndon factors} of $w$.
In the next, $\CFL(w) = (\ell_1^{n_1}, \ldots, \ell_h^{n_r})$
will be an alternative notation for the Lyndon factorization of $w$.
There is a linear time algorithm
to compute the pair $(\ell_1, n_1)$ and thus, by iteration,
the Lyndon factorization of $w$.
It is due to Fredricksen and Maiorana \cite{FM} and it is also
reported in \cite{lothaire}.
It can also be used to compute the Lyndon word in the conjugacy class of a primitive
word in linear time \cite{lothaire}.
Linear time algorithms may also be found in \cite{duval} and in the more recent paper
\cite{GGT}.

\section{Lyndon factorizations of factors of a word} \label{fact}

Let $w \in \Sigma^+$ be a word and let
$\CFL(w) = (\ell_1, \ldots, \ell_k)$ be its Lyndon factorization, $k \geq 1$.
Let $x$ be a proper factor (resp. prefix, suffix) of $w$.
We say that $x$ is a {\it simple} factor of $w$ if, for each
occurrence of $x$ as a factor of $w$, there is $j$, with $1 \leq j \leq k$,
such that $x$ is a factor of $\ell_j$.
We say that $x$ is a {\it simple} prefix (resp. suffix) of $w$ if
$x$ is a proper prefix (resp. suffix) of $\ell_1$ (resp $\ell_k$).
In this section we compare the Lyndon factorization of $w$ and that of its
non-simple factors.

The following result is a direct consequence of Theorem \ref{Lyndon-factorization}.

\begin{lemma} \label{preliminar}
Let $w \in \Sigma^+$ be a word and let
$\CFL(w) = (\ell_1, \ldots, \ell_k)$ be its Lyndon factorization.
For any $i, j$, with $1 \leq i < j \leq k$, one has
$$\CFL(\ell_{i} \cdots \ell_{j}) = (\ell_{i}, \ldots, \ell_{j}).$$
\end{lemma}

If $x$ is a non-simple factor of $w$ and $x$ does not satisfy the hypotheses of
Lemma \ref{preliminar}, then
there are $i,j$ with $1 \leq i < j \leq k$,
a suffix $\ell''_i$ of $\ell_i$ and a prefix $\ell'_j$ of $\ell_j$, with
$\ell''_i \ell'_j \not = 1$, such that
$$x = \ell''_i \ell_{i+1} \cdots \ell_{j-1} \ell'_j,$$
where it is understood that if $j = i + 1$, then $\ell_{i+1}, \ldots, \ell_{j-1} = 1$
and $\ell''_i \not = 1$, $\ell'_j \not = 1$, $\ell''_i \ell'_j \not = \ell_{i}\ell_{j}$.
We say that the sequence $\ell''_i, \ell_{i+1}, \ldots, \ell_{j-1},\ell'_j$
is {\it associated} with $x$.
The following result gives relations between $\CFL(x)$ and $\CFL(w)$.

\begin{lemma} \label{lem:compression-2}
Let $w \in \Sigma^+$ be a word and let
$\CFL(w) = (\ell_1, \ldots, \ell_k)$ be its Lyndon factorization.
Let $x$ be a non-simple factor of $w$ such that $x$ does not satisfy the hypotheses of
Lemma \ref{preliminar} and let $\ell''_i, \ell_{i+1}, \ldots, \ell_{j-1},\ell'_j$
be the sequence associated with $x$.
We have
$$\CFL(x) = (\CFL(\ell''_i), \ell_{i+1}, \ldots, \ell_{j-1}, \CFL(\ell'_j))$$
where it is understood that if $\ell''_i = 1$ (resp. $\ell'_j = 1$), then the first
term $\CFL(\ell''_i)$ (resp. last term $\CFL(\ell'_j)$) vanishes.
\end{lemma}
\begdim
Let $w, x, \ell_1, \ldots, \ell_k , \ell''_i, \ell'_j$ be as in the statement.
Set $\CFL(\ell''_i) = (m_1, \ldots , m_h)$ if $\ell''_i \not = 1$
and set $\CFL(\ell'_j) = (v_1, \ldots , v_t)$ if $\ell'_j \not = 1$.
By Theorem \ref{Lyndon-factorization}, we shall have established the lemma if
we prove the following claims
\begin{itemize}
\item[(1)]
if $\ell'_j \not = 1$ and $j > i+1$, then $\ell_{j-1} \succeq v_1$;
\item[(2)]
if $\ell''_i \not = 1$ and $j > i+1$, then $m_h \succeq \ell_{i+1}$;
\item[(3)]
if $j = i+1$, then $m_h \succeq v_1$.
\end{itemize}
We preliminary observe that
$\CFL(\ell_{j-1} \ell_j \cdots \ell_k) = (\ell_{j-1}, \ldots, \ell_{k})$
(Lemma \ref{preliminar}), hence
$\ell_{j-1}$ is the longest
prefix of $\ell_{j-1} \ell_j \cdots \ell_k$ which is a Lyndon word
(Lemma \ref{duval-prop}).

(1) If $\ell_{j-1} \prec v_1$, then
$\ell_{j-1}  v_1$ would be a Lyndon word, by Proposition \ref{propP3},
and a prefix of $\ell_{j-1} \ell_j \cdots \ell_k$,
longer than $\ell_{j-1}$, a contradiction.

(2) If $\ell''_i = \ell_i$, then $(m_1, \ldots , m_h) = (\ell_i)$ and
we are done. Otherwise, $m_h$ is a suffix of $\ell''_i$ which is a proper
nonempty suffix of $\ell_i$.
By Proposition \ref{P2}, we know that $\ell_i \prec m_h$.
If $m_h \prec \ell_{i+1}$, then $\ell_i \prec \ell_{i+1}$, in contradiction
with Eq.(\ref{LF}).

(3) Recall that in this case $x = \ell''_i \ell'_j$,
$\ell''_i \not = 1$, $\ell'_j \not = 1$, $\ell''_i \ell'_j \not = \ell_{i}\ell_{j}$.
We claim that if $m_h \prec v_1$, then $\ell_i = \ell_{j-1} \prec v_1$.
This is obvious if $\ell''_i = \ell_i$ because $(m_1, \ldots , m_h) = (\ell_i)$.
Otherwise $m_h$ is a suffix of $\ell''_i$ which is a proper
nonempty suffix of $\ell_i$.
By Proposition \ref{P2}, we know that $\ell_i \prec m_h$, thus
if $m_h \prec v_1$, then
$\ell_i = \ell_{j-1} \prec v_1$.
Hence,
$\ell_{j-1}  v_1$ would be a Lyndon word, by Proposition \ref{propP3},
and a prefix of $\ell_{j-1} \ell_j \cdots \ell_k$,
longer than $\ell_{j-1}$, a contradiction.
\enddim

Let $x,y,z,w,w' \in \Sigma^+$.
The following result, which is
a consequence of Lemma \ref{lem:compression-2}, gives relations between
the Lyndon factorizations of two overlapping words $w,w'$, i.e., such
that $w =xy$, $w'= yz$,
and the Lyndon
factorization of the overlap $y$, when $y$ is non-simple (as a suffix of
$w$ and as a prefix of $w'$).

\begin{lemma} \label{lem:compression-3}
Let $w, w' \in \Sigma^+$, let
$\CFL(w) = (\ell_1, \ldots, \ell_k)$ and $\CFL(w') = (f_1, f_{2}, \ldots, f_h)$.
If $y$ is a non-simple suffix of $w$ and a non-simple prefix of $w'$, then
there are $i,j$, with $1 \leq i < k$, $1 < j \leq h$, such that
one of the following cases holds.
\begin{itemize}
\item[(1)]
$\CFL(y) = (f_1, \ldots, f_{j-1}, \ell_{i+1}, \ldots, \ell_k)$
\item[(2)]
There exists $j'$, $1 < j' < j$ such that
$\CFL(y) = (f_1, \ldots, f_{j'-1}, \ell_{i+1}, \ldots, \ell_k)$ and
$f_{j'+r} = \ell_{i+1+r}$, for any $r$, $0 \leq r \leq j-j'-1$
\item[(3)]
There is $i'$, $i < i' < k$ such that
$\CFL(y) = (f_1, \ldots, f_{j-1}, \ell_{i'+1}, \ldots , \ell_k)$ and
$\ell_{i' - r} = f_{j - r - 1}$, for any $r$, $0 \leq r \leq i'- i -1$
\end{itemize}
\end{lemma}
\begdim
Let $w, w' \in \Sigma^+$, let $\CFL(w) = (\ell_1, \ldots, \ell_k)$ and $\CFL(w') = (f_1, f_{2}, \ldots, f_h)$.
If $y$ is a non-simple suffix of $w$ and a non-simple prefix of $w'$, then
there are $i,j$, with $1 \leq i < k$, $1 < j \leq h$, such that
$$y = \ell''_i \ell_{i+1} \cdots \ell_{k} =
f_1 \cdots f_{j-1}f'_j,$$
where $\ell''_i$ is a suffix of $\ell_i$ and
$f'_j$ is a prefix of $f_j$.
By Lemma \ref{lem:compression-2} we have
$$\CFL(y) = (\CFL(\ell''_i), \ell_{i+1}, \ldots , \ell_{k}) =
(f_1, \ldots ,f_{j-1} \CFL(f'_j)).$$
Thus the conclusion follows by
Theorem \ref{Lyndon-factorization}.
\enddim

Since Lyndon factorizations can be computed in linear time,
the above result leads to efficient measures of similarities between
words. These measures can be used to capture words that may be overlapping.

\section{Anti-Lyndon words, inverse Lyndon words and anti-prenecklaces} \label{ILW}

For the material in this section see \cite{inverseLyndon}.


\subsection{Inverse lexicographic order and anti-Lyndon words}

Inverse Lyndon words are related to the
inverse alphabetic order. Its definition is recalled below.

\begin{definition} \label{ILO}
Let $(\Sigma, <)$ be a totally ordered alphabet.
The {\rm inverse} $<_{in}$ of $<$ is defined
by
$$ \forall a, b \in \Sigma \quad b <_{in} a \Leftrightarrow a < b $$
The {\rm inverse lexicographic} or {\rm inverse alphabetic order}
on $(\Sigma^*, <)$, denoted $\prec_{in}$, is the lexicographic order
on $(\Sigma^*, <_{in})$.
\end{definition}

\begin{example}
{\rm Let $\Sigma = \{a,b,c,d \}$ with $a < b < c < d$. Then $dab \prec dabd$
and $dabda \prec dac$. We have
$d <_{in} c <_{in} b <_{in} a$. Therefore $dab \prec_{in} dabd$ and $dac \prec_{in} dabda$.}
\end{example}

The following proposition justifies the adopted terminology.

\begin{proposition} \label{P3}
Let $(\Sigma, <)$ be a totally ordered alphabet.
For all $x, y \in \Sigma^*$ such that $x \Join y$,
$$y \prec_{in} x \Leftrightarrow x \prec y .$$
Moreover, in this case $x \ll y$.
\end{proposition}

From now on, $L_{in} = L_{(\Sigma^*, <_{in})}$ denotes the set
of the Lyndon words on $\Sigma^*$ with respect to the inverse lexicographic order.
A word $w \in L_{in}$ will be named an {\it anti-Lyndon word}. Correspondingly, an
{\it anti-prenecklace} will be a prefix of an {\it anti-necklace}, which in turn will
be a necklace with respect to the inverse lexicographic order.
The following proposition characterizes $L_{in} = L_{(\Sigma^*, <_{in})}$.

\begin{proposition} \label{Lyforinv}
A word $w \in \Sigma^+$ is in $L_{in}$ if and only if $w$ is primitive and
$w \succ vu$, for each $u, v \in \Sigma^+$ such that $w = uv$.
\end{proposition}

We state below a slightly modified dual version of Proposition \ref{P2}.

\begin{proposition} \label{P4}
A word $w \in \Sigma^+$ is in $L_{in}$ if and only if $w$ is unbordered
and $w \succ v$, for each proper nonempty suffix $v$.
\end{proposition}

The following result give more precise relations between words in $L_{in}$
and their proper nonempty suffixes.

\begin{proposition} \label{P9}
If $v$ is a proper nonempty suffix of $w \in L_{in}$, then $v \ll w$.
\end{proposition}

In the following, we denote by $\CFL_{in}(w)$ the Lyndon factorization
of $w$ with respect to the inverse order $<_{in}$.

\subsection{Inverse Lyndon words and anti-prenecklaces}

\begin{definition} \label{inverse-Lyndon-word}
A word $w \in \Sigma^+$ is an inverse Lyndon word if
$s \prec w$, for each nonempty proper suffix $s$ of $w$.
\end{definition}

\begin{example}
{\rm The words $a$, $b$, $aaaaa$, $bbba$, $baaab$, $bbaba$ and $bbababbaa$
are inverse Lyndon words on $\{a,b\}$, with $a < b$.
On the contrary, $aaba$ is not an inverse Lyndon word since $aaba \prec ba$.
Analogously, $aabba \prec ba$ and thus $aabba$ is not an inverse Lyndon word.}
\end{example}

The following result is a direct consequence of Proposition \ref{P4}.

\begin{proposition} \label{P5}
A word $w \in \Sigma^+$ is an anti-Lyndon word if and only if it is
an unbordered inverse Lyndon word.
\end{proposition}

In turn, by Proposition \ref{P5} it is clear that
the set of anti-Lyndon words is a proper subset of the set of inverse Lyndon words since
there are inverse Lyndon words which are not anti-Lyndon words.
For instance consider $\Sigma = \{a,b\}$, with
$a < b$. The word $bab$ is an inverse Lyndon word but it is bordered, thus
it is not an anti-Lyndon word.

Inverse Lyndon words and anti-prenecklaces are strongly related, as the following result
shows.

\begin{proposition} \label{InvPreN}
A word $w \in \Sigma^+$ is an inverse Lyndon word if and only if $w$ is a nonempty anti-prenecklace.
\end{proposition}

The following result is a direct consequence of Proposition \ref{InvPreN}.

\begin{lemma} \label{lem:inverse-Lyndon-word-prefix}
Any nonempty prefix of an
inverse Lyndon word is an inverse Lyndon word.
\end{lemma}

\section{A canonical inverse Lyndon factorization: $\ICFL(w)$} \label{icfl}

An inverse Lyndon factorization of
a word $w \in \Sigma^+$ is a sequence $(m_1, \ldots, m_{k})$ of inverse Lyndon words
such that $m_1 \cdots m_{k} = w$ and $m_i \ll m_{i+1}$, $1 \leq i \leq k-1$.
The canonical inverse Lyndon factorization, denoted $\ICFL(w)$,
is a special inverse Lyndon factorization that
maintains the main properties of the Lyndon factorization. Its definition and properties
are based on other notions and results recalled below.

\begin{definition} \label{brepref}
Let $w \in \Sigma^+$, let $p$ be an inverse Lyndon word
which is a nonempty proper prefix
of $w = pv$.
The {\rm bounded right extension} $\overline{p}_w$ of $p$
(relatively to $w$), denoted by $\overline{p}$ when it is
understood, is a nonempty prefix of $v$ such that:
\begin{itemize}
\item[(1)]
$\overline{p}$ is an inverse Lyndon word,
\item[(2)]
$pz'$ is an inverse Lyndon word, for each proper nonempty prefix $z'$
of $\overline{p}$,
\item[(3)]
$p \overline{p}$ is not an inverse Lyndon word,
\item[(4)]
$p \ll \overline{p}$.
\end{itemize}
Moreover, we set
\begin{eqnarray*}
\Pref_{bre}(w) &=& \{(p,\overline{p}) ~|~ p \mbox{ is an inverse Lyndon word} \\
& &\mbox{which is a nonempty proper prefix of } w \}.
\end{eqnarray*}
\end{definition}

It has been proved that either
$\Pref_{bre}(w) = \emptyset$ or $\Card(\Pref_{bre}(w)) = 1$.
Moreover $\Pref_{bre}(w)$ is empty if and only if
$w$ is an inverse Lyndon word. Another useful property of $\Pref_{bre}(w)$
is recalled below.

\begin{proposition} \label{shortest}
Let $w \in \Sigma^+$ be a word which is not an inverse Lyndon word.
Let $z$ be the shortest nonempty prefix of $w$ which is not an inverse Lyndon word. Then,
\begin{itemize}
\item[(1)]
$z = p \overline{p}$, with $(p , \overline{p}) \in \Pref_{bre}(w)$.
\item[(2)]
$p = ras$ and $\overline{p} = rb$, where $r,s \in \Sigma^*$, $a,b \in \Sigma$ and
$r$ is the shortest prefix of $p \overline{p}$ such that $p \overline{p} = rasrb$,
with $a < b$.
\end{itemize}
\end{proposition}

We now give the recursive definition of $\ICFL(w)$.

\begin{definition} \label{def:ICFL}
Let $w \in \Sigma^+$.

\medskip\noindent
{\rm (Basis Step)}
If $w$ is an inverse Lyndon word,
then $\ICFL(w) = (w)$.

\medskip\noindent
{\rm (Recursive Step)}
If $w$ is not an inverse Lyndon word,
let $(p,\overline{p}) \in \Pref_{bre}(w)$ and let
$v \in \Sigma^*$ such that $w = pv$.
Let $\ICFL(v) = (m'_1, \ldots, m'_{k})$ and let
$r,s \in \Sigma^*$, $a,b \in \Sigma$ such that $p = ras$, $\overline{p} = rb$ with $a < b$.
$$\ICFL(w) = \begin{cases} (p, \ICFL(v)) & \mbox{ if } \overline{p} = rb \leq_{p} m'_1 \\
(p m'_1, m'_2, \ldots, m'_{k}) & \mbox{ if } m'_1 \leq_{p} r \end{cases}$$
\end{definition}


\section{Groupings} \label{Group}

Let $w \in \Sigma^+$. There are relations between $\ICFL(w)$ ,
the Lyndon factorization $\CFL_{in}(w)$
of $w$ with respect to the inverse order $<_{in}$ and some
special inverse Lyndon factorizations of $w$, called
{\it groupings} of $\CFL_{in}(w)$.
We first give some needed definitions
and results.

\begin{definition} \label{MaxCh}
Let $w \in \Sigma^+$, let $\CFL_{in}(w) = (\ell_1, \ldots , \ell_h)$
and let $1 \leq r < s \leq h$.
We say that $\ell_{r}, \ell_{r+1} \ldots , \ell_{s}$
is a non-increasing {\rm maximal chain for the prefix order}
in $\CFL_{in}(w)$, abbreviated $\PMCI$, if
$\ell_{r} \geq_p \ell_{r+1} \geq_p \ldots  \geq_p \ell_{s}$.
Moreover, if $r > 1$, then $\ell_{r - 1} \not \geq_p \ell_{r}$,
if $s < h$, then $\ell_{s} \not \geq_p \ell_{s +1}$.
Two $\PMCI$ $\mathcal{C}_1 = \ell_{r}, \ell_{r+1} \ldots , \ell_{s}$,
$\mathcal{C}_2 = \ell_{r'}, \ell_{r'+1} \ldots , \ell_{s'}$ are
{\rm consecutive} if $r' = s+1$ (or $r = s' +1$).
\end{definition}

The definition of a grouping of $\CFL_{in}(w)$ is given below in two steps. We first
define the grouping of a $\PMCI$. Then a grouping of $\CFL_{in}(w)$ is obtained
by changing each $\PMCI$ with one of its groupings.

\begin{definition} \label{grouping1}
Let $\ell_1, \ldots , \ell_h$ be words in $L_{in}$ such that
$\ell_i$ is a prefix of $\ell_{i-1}$, $1 < i \leq h$.
We say that $(m_1, \ldots , m_k)$ is a {\rm grouping} of $(\ell_1, \ldots , \ell_h)$
if the following conditions are satisfied.
\begin{itemize}
\item[(1)]
$m_j$ is an inverse Lyndon word,
\item[(2)]
$\ell_1 \cdots  \ell_{h} = m_1 \cdots m_k$. More precisely, there are $i_0, i_1, \ldots , i_k$,
$i_0 = 0$, $1 \leq i_j \leq h$, $i_k = h$, such that $m_j = \ell_{i_{j-1} + 1} \cdots \ell_{i_j}$,
$1 \leq j \leq k$,
\item[(3)]
$m_1 \ll \ldots \ll m_k$.
\end{itemize}
\end{definition}

We now extend Definition \ref{grouping1} to $\CFL_{in}(w)$.

\begin{definition} \label{grouping2}
Let $w \in \Sigma^+$ and let $\CFL_{in}(w) = (\ell_1, \ldots , \ell_h)$.
We say that $(m_1, \ldots , m_k)$ is a {\rm grouping} of $\CFL_{in}(w)$ if
it can be obtained by replacing
any $\PMCI$ $\mathcal{C}$ in $\CFL_{in}(w)$
by a grouping of $\mathcal{C}$.
\end{definition}

Groupings of $\CFL_{in}(w)$
are inverse Lyndon factorizations of $w$ but there are
inverse Lyndon factorizations which are not groupings.
As stated below, $\ICFL(w)$ is a grouping of
$\CFL_{in}(w)$. We first consider the special case of an inverse Lyndon word.

\begin{proposition} \label{Pr0}
Let $(\Sigma, <)$ be a totally ordered alphabet. Let $w \in \Sigma^+$ and let
$\CFL_{in}(w) = (\ell_1, \ldots , \ell_h)$.
If $w$ is an inverse Lyndon word, then either $w$ is unbordered or
$\ell_1, \ldots , \ell_{h}$ is a $\PMCI$ in $\CFL_{in}(w)$.
In both cases $\ICFL(w) = (w)$ is the unique grouping of $\CFL_{in}(w)$.
\end{proposition}

\begin{proposition} \label{Pr5}
Let $(\Sigma, <)$ be a totally ordered alphabet. For any $w \in \Sigma^+$,
$\ICFL(w)$ is a grouping of $\CFL_{in}(w)$.
\end{proposition}

\section{Borders} \label{Bord}

We recall that, given a nonempty word $w$,
a {\it border} of $w$ is a word which is both a proper prefix and a suffix
of $w$ \cite{CHL07}. The longest proper prefix of $w$ which is a suffix of
$w$ is also called {\it the border} of $w$ \cite{CHL07,lothaire}.
Thus a word $w \in \Sigma^+$ is unbordered if and only if it has a nonempty border.
Let $w \in \Sigma^+$ be a word which is not an inverse Lyndon word,
let $\ICFL(w) = (m_1, \ldots, m_{k})$.
The aim of this section is to show that any nonempty border
of $m_i$ is not a prefix of $m_{i+1}$, $1 \leq i \leq k-1$.
Some preliminary results are needed.

\begin{proposition} \label{bordo3}
Let $w \in \Sigma^+$,
let $\CFL_{in}(w) = (\ell_1, \ldots , \ell_h)$
and let $\ell_r, \ldots , \ell_s$, $1 \leq r < s \leq h$, be
a non-increasing chain for the prefix order
in $\CFL_{in}(w)$.
For any nonempty border $x$ of $y = \ell_r \cdots  \ell_s$ there is $t$,
$r \leq t < s$, such that
$x = \ell_{t+1} \cdots  \ell_s$.
Consequently, $\ell_s$ is a prefix of any nonempty border of
$\ell_r \cdots  \ell_s$.
\end{proposition}
\begdim
Let $w, \ell_1, \ldots , \ell_h, r, s$ be as in the statement.
By hypothesis, for each $t$, with $r \leq t \leq s$, the word $\ell_t$ is a prefix of $\ell_r$.
Let $x$ be a nonempty border of $y = \ell_r \cdots  \ell_s$.
If there were a nonempty proper suffix $x'$ of $\ell_t$, $r \leq t \leq s$,
such that $x = x' \ell_{t+1} \cdots  \ell_s$, then
$x'$ would be both a prefix and a nonempty proper suffix of $\ell_t$, thus a nonempty border
of $\ell_t$, in contradiction with $\ell_t$ being an anti-Lyndon word.
\enddim

\begin{lemma} \label{bordo0}
Let $w \in \Sigma^+$ be a word which is not an inverse Lyndon word,
let $\CFL_{in}(w) = (\ell_1^{n_1}, \ldots , \ell_h^{n_h})$, with
$h  > 0$, $n_1, \ldots , n_h \geq 1$.
For all $z \in \Sigma^+$ and $b \in \Sigma$
such that $z$ is an anti-prenecklace, $zb$ is not an anti-prenecklace
and $zb$ is a prefix of $w$, there is an integer $g$ such that
$$zb = (u_1v_1)^{n_1} \cdots (u_gv_g)^{n_g} u_gb,$$
where $u_jv_j = u_ja_jv'_j = \ell_j, 1 \leq j \leq g, a_j < b$
and $u_gb$ is an anti-prenecklace.
\end{lemma}
\begdim
We prove the statement by induction on $|w|$.
If $|w| = 1$, then $w$ is an inverse Lyndon word and we are done.
Hence assume $|w| > 1$.
If $w$ is an inverse Lyndon word,
then again the proof is ended.
Therefore, assume that $w$ is not an inverse Lyndon word.
Let $\CFL_{in}(w) = (\ell_1^{n_1}, \ldots , \ell_h^{n_h})$, with
$h  > 0$, $n_1, \ldots , n_h \geq 1$.

Let $z \in \Sigma^+$, $b \in \Sigma$ be
such that $z$ is an anti-prenecklace, $zb$ is not an anti-prenecklace
and $zb$ is a prefix of $w$.
By Theorem \ref{FundamentalPreneck} and Corollary \ref{FundamentalPreneckCor},
there are words $u,v, v' \in \Sigma^*$, $a \in \Sigma$, with $a < b$, and an integer $k \geq 1$,
such that
$zb = (uv)^kub$, $v = av'$ and where $uv$ is the longest anti-Lyndon prefix of $zb$.

We claim that $uv$ is also the longest anti-Lyndon prefix of $w$.
Indeed, if $y$ is a prefix of $w$ such that $|y| > |zb|$, then $y = zbz' = (uav')^kubz'$, with
$z' \in \Sigma^*$. Thus, $y \ll ubz'$ and $y$ is not an anti-Lyndon word.
Consequently, by Lemma \ref{duval-prop}, we have $uv = \ell_1$. Moreover, $k = n_1$ because $ub$ is not a prefix of
$\ell_1$.

If $ub$ is an anti-prenecklace the proof is ended. Otherwise,
let $w' \in \Sigma^*$ be such that $w = \ell_1^{n_1} w'$. We have $0 < |w'| < |w|$ since
$ub$ is a prefix of $w'$ and $\ell_1 \not = 1$.
By Theorem \ref{Lyndon-factorization},
we have $\CFL_{in}(w') = (\ell_2^{n_2}, \ldots , \ell_h^{n_h})$.
The word $u$ an anti-prenecklace whereas $ub$ is not an anti-prenecklace.
By induction hypothesis there is an integer $g$ such that
$$ub = (u_2v_2)^{n_2} \cdots (u_gv_g)^{n_g} u_gb,$$
where $u_jv_j = u_ja_jv'_j = \ell_j, 2 \leq j \leq g, a_j < b$
and $u_gb$ is an anti-prenecklace.
Thus, there is an integer $g$ such that
$$zb = (u_1v_1)^{n_1} \cdots (u_gv_g)^{n_g} u_gb,$$
where $u_1 = u$, $v_1 = v$, $v'_1 = v'$, $a_1 = a$,
$u_jv_j = u_ja_jv'_j = \ell_j, 1 \leq j \leq g, a_j < b$
and $u_gb$ is an anti-prenecklace.
\enddim

\begin{proposition} \label{bordo1}
Let $w \in \Sigma^+$ be a word which is not an inverse Lyndon word,
let $(p, \bar{p}) \in \Pref_{bre}(w)$ and let
$\ICFL(w) = (m_1, \ldots , m_k)$. Let
$\CFL_{in}(w) = (\ell_1^{n_1}, \ldots , \ell_h^{n_h})$, with
$h  > 0$, $n_1, \ldots , n_h \geq 1$ and let
$\ell_1^{n_1}, \ldots , \ell_q^{n_q}$ be a
$\PMCI$ in $\CFL_{in}(w)$, $1 \leq q \leq h$.
Then the following properties hold.
\begin{itemize}
\item[(1)]
$p = \ell_1^{n_1} \cdots \ell_{g}^{n_{g}}$,
for some $g$, $1 \leq g \leq q$.
\item[(2)]
$\ell_{g} = u_g v_g = u_g a_g v'_g$, $\bar{p} = u_g b$, $a_g < b$.
\end{itemize}
\end{proposition}
\begdim
Let $w \in \Sigma^+$ be a word which is not an inverse Lyndon word,
let $(p, \bar{p}) \in \Pref_{bre}(w)$.
Let $\CFL_{in}(w) = (\ell_1^{n_1}, \ldots , \ell_h^{n_h})$, with
$h  > 0$, $n_1, \ldots , n_h \geq 1$ and let
$\ell_1^{n_1}, \ldots , \ell_q^{n_q}$ be a
$\PMCI$ in $\CFL_{in}(w)$, $1 \leq q \leq h$.

By Proposition \ref{InvPreN}, the word $p \bar{p}$ is not an anti-prenecklace but
its longest proper prefix is an anti-prenecklace.
Thus, by Lemma \ref{bordo0} there is an integer $g$ such that
$$p \bar{p} = (u_1v_1)^{n_1} \cdots (u_gv_g)^{n_g} u_gb,$$
where $u_jv_j = u_ja_jv'_j = \ell_j, 1 \leq j \leq g, a_j < b$
and $u_gb$ is an anti-prenecklace.
Let
$$\beta = (u_1v_1)^{n_1}(u_2v_2)^{n_2} \cdots (u_gv_g)^{n_g},
\quad \beta' = \beta u_g.$$
By Definition \ref{brepref}, the words $\beta'$ and $\beta$ are inverse Lyndon words, therefore
$g \leq q$ (otherwise $\ell_q$ would be a prefix of $\beta$ and there would be a word $z'$ such that
$\ell_{q+1}z'$ is a suffix of $\beta$, a contradiction
since $\beta$ is an inverse Lyndon word and $\ell_q \ll \ell_{q+1}$).
Moreover, $\beta \ll u_gb$.

Let $r, s \in \Sigma^*$, $a', b \in \Sigma$ be such that $p = ra's$, $\bar{p} = rb$, $a' < b$.
Then $p \bar{p} = \beta u_gb = r a's rb$. By Proposition \ref{shortest}, $r$ is a suffix of
$u_g$. Consequently, $\ell_g = u_gv_g$ and $u_g$ are prefixes of $p$.
Moreover, we know that $u_g$ and $u_gb$ are both anti-prenecklaces.
Thus, by Proposition \ref{preprime}, Theorem \ref{FundamentalPreneck}
and Corollary \ref{FundamentalPreneckCor},
there are $x, y \in \Sigma^*$, an integer $t \geq 1$, $c \in \Sigma$ such that
$xy$ is an anti-Lyndon word, $u_g = (xy)^t x$,
$y = cy'$ with $c \geq b$.

The words $\ell_{g+1}$ and $u_gb = (xy)^t xb$ are both prefixes of the
same word $\gamma$, hence they are comparable for the prefix order. Since $\ell_{g+1}$
is the longest anti-Lyndon prefix of
$\gamma$, we have $|\ell_{g+1}| \geq |xy|$ and since $\ell_{g+1}$ is unbordered,
either $\ell_{g+1} = xy$ is a prefix of $\ell_{g}$ and $g+1 \leq q$,
or the word $u_gb = (xy)^t xb$ is a prefix of $\ell_{g+1}$.
By Proposition \ref{Pr5},
the first case holds, otherwise $m_1$ would not be a product
of anti-Lyndon words because $m_1$ is a prefix of $\beta u_g$.

If $r = u_g$, then $p = \beta$ and the proof is ended. By contradiction, assume that
$r$ is a proper suffix of $u_g$. Therefore $r$ is a border of $u_g$ because $r$ is a prefix of
$p$ and $u_g$ is nonempty.
Of course $r \not = x$ because $u_g$ starts with $ra'$ and also with $xc$,
with $c \geq b > a'$.
If $r$ would be shorter than $x$, then $r$ would be a border of $x$.
This is impossible because $rcy'(xy)^{t-1}x$ would be a suffix of
the inverse Lyndon word $u_g$ and $u_g$ starts with $ra'$,
with $c \geq b > a'$. Thus $|r| > |x| \geq 0$.
Since $r$ is a nonempty border of $u_g = (xy)^t x$ and $|r| > |x| \geq 0$,
one of the following three cases holds:
\begin{eqnarray}
r & = & (xy)^{t'}x, \quad 0 < t' < t   \label{eq1} \\
r & = & y_1(xy)^{t'}x, \quad y_1 \mbox{ nonempty suffix of } y,  \quad 0 \leq  t' < t  \label{eq2} \\
r & = & x_1(yx)^{t'}, \quad x_1 \mbox{ nonempty suffix of } x,  \quad 0 < t' \leq t \label{eq3}
\end{eqnarray}
Assume that Eq. (\ref{eq1}) holds. Then $p$ starts with $ra' = (xy)^{t'}xa'$, $a' < b$,
and $p$ also starts with $u_g = (xy)^tx$. Since $t' < t$, the letter $a'$ should be the first
letter of $y = cy'$, $c \geq b > a'$. Therefore, Eq. (\ref{eq1}) cannot hold.

Assume that Eq. (\ref{eq2}) holds. Therefore $y_1 =y$, otherwise $y_1x$ would be a proper prefix
of $xy$, hence a nonempty border of $xy$, which is impossible since $xy$ is an anti-Lyndon word.
Moreover $x = 1$, otherwise $yx = xy$ and $xy$ would not be primitive (Proposition \ref{nocode}),
which is impossible since $xy$ is an anti-Lyndon word.
As above, $p$ starts with $ra' = y^{t'}a'$, $a' < b$,
and $p$ also starts with $u_g = y^t$. Since $t' < t$, the letter $a'$ should be the first
letter of $y = cy'$, $c \geq b > a'$. Therefore, Eq. (\ref{eq2}) cannot hold.

Finally, assume that Eq. (\ref{eq3}) holds.
If $x_1 \not = x$, then $x_1y$ would be both a proper nonempty suffix
and a prefix of $xy$, hence a nonempty border of $xy$,
which is impossible since $xy$ is an anti-Lyndon word.
Therefore $x_1 = x$. If $t' < t$, then $r$ satisfies Eq. (\ref{eq1})
and we proved that this is impossible.
Thus $t' = t$, which implies $r = u_g$, a contradiction.
\enddim

\begin{proposition}\label{bordo-bre-paola}
Let $w \in \Sigma^+$ be a word which is not an inverse Lyndon word and
let $(p, \bar{p}) \in \Pref_{bre}(w)$.
For each nonempty border $z$ of $p$,
one has that $z$ and $\overline{p}$ are incomparable for the prefix order.
\end{proposition}
\begdim
Let $w \in \Sigma^+$ be a word which is not an inverse Lyndon word and
let $(p, \bar{p}) \in \Pref_{bre}(w)$.
By Proposition \ref{shortest}, there are $r,s \in \Sigma^*$, $a,b \in \Sigma$,
with $a < b$, such that
$p = ras$ and $\overline{p} = rb$.
Let $\CFL_{in}(w) = (\ell_1^{n_1}, \ldots , \ell_h^{n_h})$, with
$h  > 0$, $n_1, \ldots , n_h \geq 1$ and let
$\ell_1^{n_1}, \ldots , \ell_q^{n_q}$ be a
$\PMCI$ in $\CFL_{in}(w)$, $1 \leq q \leq h$.

Let $z$ be a nonempty border of $p$. Of course $\overline{p}$ cannot be a prefix
of $z$ because $\overline{p}$ is not a prefix of $p$.
By contradiction, suppose that $z$ is a prefix of $\overline{p}$.
By Proposition \ref{bordo1}, there is $g$, $1 \leq g \leq q$ such that
$p = \ell_1^{n_1} \cdots \ell_{g}^{n_{g}}$
and $\ell_{g} = u_g v_g = u_g a_g v'_g$, $\bar{p} = u_g b$, $a_g < b$.

By Proposition \ref{bordo3}, $\ell_{g}$ is a prefix of any nonempty border of $p$, hence
$\ell_{g}$ is a prefix of $z$. Moreover $z$ is a prefix of $\overline{p}$,
thus $\ell_g=u_ga_gv'_g$ would be a prefix of $\overline{p}=u_gb$.
This is impossible because $a_g < b$.
\enddim

\begin{proposition} \label{bordo2}
Let $w \in \Sigma^+$ be a word which is not an inverse Lyndon word and
let $\ICFL(w) = (m_1, \ldots , m_k)$.
If $z$ is a nonempty border of $m_1$, then $z$ is not a prefix of $m_2$.
\end{proposition}
\begdim
Let $w \in \Sigma^+$ and
let $\ICFL(w) = (m_1, \ldots , m_k)$.
We prove the statement by induction on $|w|$.

If $|w| = 1$, then $w$ is an inverse Lyndon word and we are done.
Hence assume $|w| > 1$.
If $w$ is an inverse Lyndon word,
then again the proof is ended.
Therefore, assume that $w$ is not an inverse Lyndon word.
Let $(p, \bar{p}) \in \Pref_{bre}(w)$ and let
$r,s \in \Sigma^*$, $a,b \in \Sigma$ be such that
$p = ras$, $\overline{p} = rb$, $a < b$.
Let $v \in \Sigma^*$ be such that $w = m_1v$.
Of course $0 < |v| < |w|$ because $w$ is not an inverse Lyndon word.
Let $\ICFL(v) = (m'_1, \ldots , m'_{k'})$.
By Definition \ref{def:ICFL}, one of the following two cases holds
\begin{itemize}
\item[(1)]
$m_1 = p$ if $\overline{p}$ is a prefix of $m'_1 = m_2$
\item[(2)]
$m_1 = pm'_1$, $m_2 = m'_2, \ldots , m_k = m'_{k}$, $k'=k$,  if $m'_1$ is a prefix of $r$.
\end{itemize}

Let $z$ be a nonempty border of $m_1$.
In case (1), if $z$ would be a prefix of $m_2$, then
$z$ and $\overline{p}$ would be comparable
for the prefix order, in contradiction with Proposition \ref{bordo-bre-paola}.

In case (2), $m'_1$ is a prefix of $m_1$.
By contradiction, suppose that $z$ is a prefix of $m_2$.
We have either $|z| \geq |m'_1|$ or $|z| < |m'_1|$.
If $|z| \geq |m'_1|$, then
$m'_1$ would be a prefix of $z$ and thus of $m_2 = m'_2$, in contradiction with $m'_1 \ll m'_2$.
If $|z| < |m'_1|$, then $z$ would be a suffix of $m'_1$,
hence $z$ would be a nonempty border of $m'_1$.
Thus a nonempty border of $m'_1$ would be
a prefix of $m_2 = m'_2$, in contradiction with the induction hypothesis.
\enddim

\begin{proposition} \label{bordo4}
Let $w \in \Sigma^+$ be a word which is not an inverse Lyndon word and
let $\ICFL(w) = (m_1, \ldots , m_k)$.
If $z$ is a nonempty border of $m_i$, then $z$ is not a prefix of $m_{i+1}$,
$1 \leq i \leq k-1$.
\end{proposition}
\begdim
Let $w \in \Sigma^+$ and
let $\ICFL(w) = (m_1, \ldots , m_k)$.
We prove the statement by induction on $|w|$.

If $|w| = 1$, then $w$ is an inverse Lyndon word and we are done.
Hence assume $|w| > 1$.
If $w$ is an inverse Lyndon word,
then again the proof is ended.
Therefore, assume that $w$ is not an inverse Lyndon word.
Let $(p, \bar{p}) \in \Pref_{bre}(w)$ and let
$r,s \in \Sigma^*$, $a,b \in \Sigma$ be such that
$p = ras$, $\overline{p} = rb$, $a < b$.
Let $v \in \Sigma^*$ be such that $w = m_1v$.
Of course $0 < |v| < |w|$ because $w$ is not an inverse Lyndon word.
Let $\ICFL(v) = (m'_1, \ldots , m'_{k'})$.
By Definition \ref{def:ICFL}, one of the following two cases holds
\begin{itemize}
\item[(1)]
$m_1 = p$, $m_{i} = m'_{i - 1}$, $1 < i \leq k = k' + 1$, if $\overline{p} \leq_{p} m'_1$
\item[(2)]
$m_1 = pm'_1$, $m_{i} = m'_{i}$, $1 \leq i \leq k = k'$, if $m'_1 \leq_{p} r$.
\end{itemize}

If $z$ is a nonempty border of $m_1$, then $z$ is not a prefix of $m_{2}$,
by Proposition \ref{bordo2}.
Thus assume that $z$ is a nonempty border of $m_i$, $1 < i \leq k-1$.
In case (1), $z$ is a nonempty border of $m'_{i - 1}$, hence,
by induction hypothesis, $z$ is not a prefix of $m'_{i} = m_{i+1}$.
Analogously, in case (2), $z$ is a nonempty border of $m'_i$, therefore,
by induction hypothesis, $z$ is not a prefix of $m'_{i+1} = m_{i+1}$.

\section{Suffixes compatibility} \label{SC}

In this section we use the same notation and terminology as in
\cite{restivo-sorting,restivo-sorting-2014}, where the authors
found interesting relations between the sorting of the suffixes of a word $w$
and that of its factors. Here we prove a similar property when $\ICFL(w)$ is considered.

Let $w, x, u, y \in \Sigma^*$, and let $u$ be a nonempty factor of $w = xuy$.
Let $first(u)$ and $last(u)$ denote
the position of the first and the last symbol of $u$ in $w$, respectively.
If $w=a_1 \cdots a_n$, $a_i \in \Sigma$, $1 \leq i \leq j \leq n$,
then we also set $w[i, j]=a_i \cdots a_j$.
A \textit{local suffix} of $w$ is a suffix of a factor of $w$, specifically
$suf_u(i) = w[i, last(u)]$ denotes the \textit{local suffix} of $w$
at the position $i$ with respect to $u$, $i \geq first(u)$.
The corresponding \textit{global suffix} $suf_u(i)y$ of
$w$ at the position $i$ is denoted by
$suf_w(i) = w[i, last(w)]$ (or simply $suf(i)$ when it is understood).
We say that $suf_u(i)y$ is {\it associated} with $suf_u(i)$.

\begin{definition} \cite{restivo-sorting,restivo-sorting-2014} \label{def-sorting}
Let $w \in \Sigma^+$ and let $u$ be a nonempty factor of $w$.
We say that the sorting of the nonempty local suffixes of
$w$ with respect to $u$ is {\rm compatible}
with the sorting of the corresponding nonempty global suffixes of $w$
if for all $i,j$ with $first(u) \leq i < j \leq last(u)$,
$$suf_u(i) \prec suf_u(j) \Longleftrightarrow suf(i) \prec suf(j).$$
\end{definition}

The following result has been proved in \cite{restivo-sorting,restivo-sorting-2014}.

\begin{theorem} \label{teo-sorting}
Let $w \in \Sigma^+$ and let $\CFL(w) = (\ell_1, \ldots, \ell_h)$ be
its Lyndon factorization.
Then, for any $i,g$, $1 \leq i \leq g \leq h$,
the sorting of the nonempty local suffixes of
$w$ with respect to $u = \ell_i \cdots \ell_g$
is compatible with the sorting of the corresponding nonempty global suffixes of $w$.
\end{theorem}

In \cite{inverseLyndon} the same compatibility property as in Theorem \ref{teo-sorting}
has been proved
for the sorting of the nonempty suffixes of a word $w$ with respect to $\prec_{in}$,
when we replace $\CFL(w)$ with $\ICFL(w)$.

\begin{proposition} \label{teo-sorting-inverseBIS}
Let $w$ be a word and let $\ICFL(w) = (m_1, \ldots , m_k)$.
Then, for any $i,h$, $1 \leq i \leq h \leq k$,
the sorting with respect to $\prec_{in}$
of the nonempty local suffixes of
$w$ with respect to $u = m_i \cdots m_h$
is compatible with the sorting with respect to $\prec_{in}$
of the corresponding nonempty global suffixes of $w$.
\end{proposition}

The following result proves another compatibility property
for the sorting of the nonempty suffixes of a word $w$ with respect to $\prec$,
when we replace $\CFL(w)$ with $\ICFL(w)$.

\begin{proposition} \label{compatibility-order-inverse-factorization}
Let $w \in \Sigma^+$ be a word which is not an inverse Lyndon word and
let $\ICFL(w)=(m_1, \ldots, m_k)$.
Let $u = m_im_{i+1} \cdots m_h$ with $1 \leq i < h \leq k$.
Assume that $suf_u(j_1) \prec suf_u(j_2)$, where
$first(u) \leq j_1 \leq last(u)$,
$first(u) \leq j_2 \leq last(u)$, $j_1 \not = j_2$.

If $suf_u(j_1)$ is a proper prefix of $suf_u(j_2)$ and
$h < k$ then $suf(j_2) \prec suf(j_1)$, otherwise $suf(j_1) \prec suf(j_2)$.
\end{proposition}
\begdim
Let $w \in \Sigma^+$ be a word which is not an inverse Lyndon word and
let $\ICFL(w)=(m_1, \ldots, m_k)$.
Let $u = m_im_{i+1} \cdots m_h$ with $1 \leq i < h \leq k$.
Assume that $suf_u(j_1) \prec suf_u(j_2)$, where
$first(u) \leq j_1 \leq last(u)$,
$first(u) \leq j_2 \leq last(u)$.

If $h = k$, then $suf(j_1) = suf_u(j_1) \prec suf_u(j_2) = suf(j_2)$
and we are done.
Thus assume $h < k$.
If $suf_u(j_1)$ is not a proper prefix of $suf_u(j_2)$, then
$suf_u(j_1) \ll suf_u(j_2)$. Hence, by item (2) in Lemma \ref{proplexord}, we have
$suf(j_1) \ll suf(j_2)$ and we are done again.

Therefore, assume that $suf_u(j_1)$ is a proper prefix of $suf_u(j_2)$.
Thus $j_2 < j_1$ because $|suf_u(j_1)| <  |suf_u(j_1)|$.
Set $x = suf_u(j_1)=w[j_1, last(m_h)]$ and $y = w[j_2, j_2 + |x| - 1]$.
We have $x = y$ because $x,y$ are prefixes of $suf_u(j_2)$
with the same length.
Let $g$ be the minimum integer such that
$j_2 + |x| \leq last(m_g)$, $g \leq h < k$, and let
$z = w[j_2 + |x|, last(m_g)]$.
Therefore,
$$suf(j_2) = xzm_{g + 1} \cdots m_k, \quad suf(j_1) = xm_{h + 1} \cdots m_k$$
and we distinguish two cases:
\begin{itemize}
\item[(1)]
$z = 1$
\item[(2)]
$z \not = 1$
\end{itemize}
(Case (1)) If $z = 1$, then $g < h$ because $j_2 \not = j_1$ and thus $suf(j_2) \not = suf(j_1)$.
Therefore
$$suf(j_2) = xm_{g + 1} \cdots m_k \ll xm_{h + 1} \cdots m_k = suf(j_1)$$
(Case (2)) Assume $z \not = 1$. If $z = m_g$, we apply the above argument again
and we obtain
$$suf(j_2) = xm_gm_{g + 1} \cdots m_k \ll xm_{h + 1} \cdots m_k = suf(j_1)$$
Thus assume that $z$ is a nonempty proper suffix of $m_g$.
Hence $z \prec m_g$ and we have one of the following two cases.
\begin{itemize}
\item[(2a)]
$z \ll m_g$
\item[(2b)]
$z <_p m_g$
\end{itemize}
(Case (2a)) If $z \ll m_g$, then we have
$$suf(j_2) = xzm_{g + 1} \cdots m_k \ll xm_gm_{g + 1} \cdots m_k \ll xm_{h + 1} \cdots m_k = suf(j_1)$$
(Case (2b))
Let $r, s \in \Sigma^*$, $a, b \in \Sigma$ be such that $m_g = ras$, $m_{g+1} = rbt$, $a < b$.
Assume that $z <_p m_g$. Since $z$ is
also a nonempty proper suffix of $m_g$, we have that $z$ is a border of $m_g$.
Then, by Proposition \ref{bordo4}, $z$ cannot be a prefix of $m_{g+1}$, hence
there is a prefix $s'$ of $s$ such that $z = ras'$.
Therefore we have
$$suf(j_2) = xzm_{g + 1} \cdots m_k \ll xm_{g + 1} \cdots m_k \preceq xm_{h + 1} \cdots m_k = suf(j_1)$$
and the proof is complete.
\enddim

\begin{example}\label{ex:case1-1}
{\rm Let $w = a^{12}bbab \in \{a,b \}^+$
with $a < b$.
We have $\ICFL(w) = (m_1, m_2) = (a^{12},bbab)$.
Let $u = m_1 = a^{12}$.
Consider $suf_u(4) = a^9$ and $suf_{u}(12) = a$.
We have $suf_{u}(12) = a \prec a^9 = suf_u(4)$.
We are in the first case of
Lemma \ref{compatibility-order-inverse-factorization}
and then
$suf(4) = a^9bbab \prec abbab = suf(12)$.}
\end{example}

\begin{example} \label{ex:case1-2}
{\rm Let $w = dabadabdabdadac \in \{a,b,c,d \}^+$
with $a < b < c < d$.
We have $\ICFL(w)= (m_1,m_2,m_3) = (daba, dabdab, dadac)$.
Let $u = m_2$. Consider $suf_{m_2}(8)= dab$ and $suf_{m_2}(5)= dabdab$.
We have $suf_{m_2}(8)= dab \prec suf_{m_2}(5)= dabdab = (dab)^2$.
We are in the first case of Lemma \ref{compatibility-order-inverse-factorization}
and then
$suf(5) = dabdabdadac \prec suf(8) = dabdadc$.

Consider now $suf_{m_2}(9)= ab \prec suf_{m_2}(8)= dab$.
Since $suf_{m_2}(9)$ is not a proper prefix of $suf_{m_2}(8))$,
we are in the second case of Lemma \ref{compatibility-order-inverse-factorization}
and we have $suf(9)=abdadac \prec suf(8)=dabdadac$.}
\end{example}

\section{Sorting Suffixes via $\ICFL$} \label{ss}

Let $w \in \Sigma^+$ be a word which is not an inverse Lyndon word.
The aim of this section is to define an integer related
to $\ICFL(w)$ and then to prove that it is an upper bound to the lengths
$\LCP(x,y)$ of the {\it longest common prefix} $\lcp(x,y)$
of two factors $x,y$ of $w$.
Some preliminary results are needed and proved below.

\subsection{Technical Results}

\begin{lemma} \label{factor}
Let $w \in \Sigma^+$ be a word which is not an inverse Lyndon word.
Let $\ICFL(w)=(m_1, \ldots, m_k)$. Then $m_i \not \in \Fact(m_1 \cdots m_{i-1})$,
for each $1 < i \leq k$.
\end{lemma}
\begdim
Let $w \in \Sigma^+$ be a word which is not an inverse Lyndon word.
Let $\ICFL(w)=(m_1, \ldots, m_k)$.
Suppose the lemma were false. Then there would be $i$, $1 < i \leq k$,
such that $m_i \in \Fact(m_1 \cdots m_{i-1})$.
Thus one of the following three cases holds.
\begin{itemize}
\item[(1)] There are an integer $j$, $1 \leq j < i$, and $x,y \in \Sigma^*$
such that $m_j = xm_iy$
\item[(2)] There is an integer $j$, $1 \leq j < i$, such that $m_j$ is a prefix of $m_i$.
\item[(3)] There are integers $j, h$, $1 \leq j < i$, $h \geq 0$, a proper nonempty suffix $x$ of $m_j$,
and a proper prefix $y$ of $m_{j+h +1}$ such that $m_i = xm_{j+1} \cdots m_{j+h}y$, where it is understood
that $m_{j+1} \cdots m_{j+h} = 1$ for $h = 0$.
\end{itemize}
Assume that case (1) holds.
If $x = 1$, then $m_i \preceq m_j \ll m_i$ which contradicts
Lemma \ref{proplexordbis}.
Otherwise, $m_iy$ is a proper suffix of $m_j$, hence
$m_iy \preceq m_j \ll m_i$. Therefore $m_iy \preceq m_j \ll m_iy$
(Lemma \ref{proplexord}) which is impossible, once again
by Lemma \ref{proplexordbis}.
Case (2) leads also to a contradiction since in this case we would have
$m_{j} <_{p} m_i$ whereas $m_{j} \ll m_i$.

Assume that case (3) holds. We know that
$x \preceq m_j$. If $x \ll m_j$, then $m_i \ll m_j$
(Lemma \ref{proplexord}) which is impossible since $m_j \ll m_i$ and then
$m_j \ll m_i \ll m_j$, in contradiction with Lemma \ref{proplexordbis}.
Thus $x$ is a proper prefix, thus a border of
$m_j$. By Proposition \ref{bordo4}, $x$ is not a prefix of $m_{j+1}$.
Thus $j+1 < i$ and there are
$r, s, t \in \Sigma^*$,
$a, b \in \Sigma$ be such that $m_{j+1} = ras$, $m_i = rbt$, $a < b$.
The words $x, r$ are comparable for the prefix order and $x$ is not
a prefix of $m_{j+1}$. Therefore there is $t' \in \Sigma^*$
such that $x = rbt'$. Consequently, $m_{j+1} \ll x$, hence
$m_{j+1} \ll m_j$ (Lemma \ref{proplexord}). Since
$m_j \ll m_{j+1}$, we would have $m_{j+1} \ll m_j \ll m_{j+1}$,
once again in contradiction with Lemma \ref{proplexordbis}.
\enddim

\begin{lemma}\label{preliminary-zero}
Let $w \in \Sigma^+$ be a word which is not an inverse Lyndon word and
let $\ICFL(w) = (m_1, \ldots, m_k)$. Let $i, h, j$ be integers
such that $1 \leq i < h < j \leq k$.
Let $r_i, s_i, t_i, r_h, s_h, t_h  \in \Sigma^*$, $a_i, b_i, a_h, b_h \in \Sigma$ be such that
$m_i = r_ia_is_i$,
$m_h = r_ha_hs_h$, $m_j = r_ib_it_i = r_hb_h t_h$, $a_i < b_i$, $a_h < b_h$.
Then, the word $r_i$ is a prefix of $r_h$.
\end{lemma}
\begdim
Let $w \in \Sigma^+$ be a word which is not an inverse Lyndon word and
let $\ICFL(w) = (m_1, \ldots, m_k)$. Let $i, h, j$ be integers
such that $1 \leq i < h < j \leq k$.
Let $r_i, s_i, t_i, r_h, s_h, t_h  \in \Sigma^*$, $a_i, b_i, a_h, b_h \in \Sigma$ be such that
$m_i = r_ia_is_i$,
$m_h = r_ha_hs_h$, $m_j = r_ib_it_i = r_hb_h t_h$, $a_i < b_i$, $a_h < b_h$.
The words $r_i$ and $r_h$ are comparable for the prefix order.
If $r_h$ would be a proper prefix of $r_i$, then
$r_hb_h$ were a prefix of $r_i$. Thus there would be
$u \in \Sigma^*$ such that
$r_i = r_hb_hu$, and consequently
$m_h = r_ha_hs_h \ll r_hb_hua_is_i = m_i \ll m_h$,
which is impossible (Lemma \ref{proplexordbis}).
Thus $r_i$ is a prefix of $r_h$.
\enddim

\begin{corollary}\label{corollariopreliminary-zero}
Let $w \in \Sigma^+$ be a word which is not an inverse Lyndon word and
let $\ICFL(w) = (m_1, \ldots, m_k)$. Let $i, h, j$ be integers
such that $1 \leq i < h < j \leq k$.
Let $r, s, t \in \Sigma^*$, be such that
$m_i = rs$ and
$m_j = rt$.
Then, the word $r$ is a prefix of $m_h$.
\end{corollary}
\begdim
Let $w \in \Sigma^+$ be a word which is not an inverse Lyndon word and
let $\ICFL(w) = (m_1, \ldots, m_k)$. Let $i, h, j$ be integers
such that $1 \leq i < h < j \leq k$.
Let $r, s, t \in \Sigma^*$, be such that
$m_i = rs$ and
$m_j = rt$.
Let $r_i, s_i, t_i, r_h, s_h, t_h  \in \Sigma^*$, $a_i, b_i, a_h, b_h \in \Sigma$ be such that
$m_i = r_ia_is_i$,
$m_h = r_ha_hs_h$, $m_j = r_ib_it_i = r_hb_h t_h$, $a_i < b_i$, $a_h < b_h$.
Of course $r$ is a prefix of $r_i$, because $r \in \Pref(m_i) \cap \Pref(m_j)$.
Thus $r$ is a prefix of $r_h$, by Lemma \ref{preliminary-zero},
hence $r \in \Pref(m_h)$.
\enddim

Let $w \in \Sigma^+$ be a word which is not an inverse Lyndon word and
let $\ICFL(w) = (m_1, \ldots, m_k)$. Let $i$ be an integer
such that $1 < i \leq k$.
Let $r_h, s_h, t_h  \in \Sigma^*$, $a_h, b_h \in \Sigma$ be such that
$m_h = r_ha_hs_h$,
$m_i = r_hb_ht_h$, $a_h < b_h$, $1 < h \leq i-1$.
The following strengthening of Lemma \ref{factor} is proved below:
$r_{i-1}b_{i-1} \not \in \Fact(m_1 \cdots m_{i-1})$ (Lemma \ref{Lemma3}).
We have divided the proof of this result into a sequence of lemmas.
We first prove that $r_{i-1}b_{i-1} \not \in \Fact(m_h)$,
$1 \leq h \leq i-1$ (Lemma \ref{Lemma1}). Then, we prove that
$r_{i-1}b_{i-1} \not \in \Fact(m_hm_{h+1})$,
$1 \leq h < i-1$ (Lemma \ref{Lemma2}).
Finally, we prove Lemma \ref{Lemma3}.

\begin{lemma} \label{Lemma1}
Let $w \in \Sigma^+$ be a word which is not an inverse Lyndon word and
let $\ICFL(w) = (m_1, \ldots, m_k)$. Let $i$ be an integer
such that $1 < i \leq k$.
Let $r_h, s_h, t_h  \in \Sigma^*$, $a_h, b_h \in \Sigma$ be such that
$m_h = r_ha_hs_h$,
$m_i = r_hb_ht_h$, $a_h < b_h$, $1 \leq h \leq i-1$.
Then, for each $h$, with $1 \leq h \leq i-1$,
we have $r_{i-1}b_{i-1} \not \in \Fact(m_h)$.
\end{lemma}
\begdim
Let $w \in \Sigma^+$ be a word which is not an inverse Lyndon word and
let $\ICFL(w) = (m_1, \ldots, m_k)$. Let $i$ be an integer
such that $1 < i \leq k$.
Let $r_h, s_h, t_h  \in \Sigma^*$, $a_h, b_h \in \Sigma$ be
as in the statement.

Suppose, contrary to our claim, that there exists
$h$, with $1 \leq h \leq i-1$, such that
$r_{i-1}b_{i-1} \in \Fact(m_h)$.
Therefore, there are $u, v \in \Sigma^*$
such that $m_h = u r_{i-1}b_{i-1} v$.
If $r_{i-1}b_{i-1}$ were a prefix of $m_h$, then necessarily
$h < i-1$, because $m_{i-1}$ starts with $r_{i-1}a_{i-1}$. Thus,
by $r_{i-1}a_{i-1} \ll r_{i-1}b_{i-1}$ we would
have $m_{i-1} \ll m_h$, with $h < i-1$, which is impossible.
Hence, $r_{i-1}b_{i-1} v$ is a proper nonempty suffix of $m_h$.
Since $r_{i-1}b_{i-1} v$ is a proper nonempty suffix of $m_h$ and
$r_{i-1}b_{i-1} v \not \in \Pref(m_h)$,
we have
$r_{i-1}b_{i-1} v \ll m_h$.
By definition, there are $r,s,t \in \Sigma^*$, $a,b \in \Sigma$, such that
$r_{i-1}b_{i-1} v = ras$, $m_h =rbt$, $a < b$.
The words $r_{i-1}$ and $r$ are comparable for the prefix order.
Moreover, $r_{i-1}$ cannot be a proper prefix of $r$ because
$r_{i-1}b_{i-1} \not \in \Pref(m_h)$.
Hence, $r$ is a prefix of $r_{i-1}$, thus
$ra$ is a prefix of $r_{i-1}b_{i-1}$.
As a consequence we have
$r_{i-1}b_{i-1} \ll m_h$ which yields
$m_i \ll m_h$, with $h < i$, once again a contradiction.
\enddim

\begin{lemma} \label{Lemma2}
Let $w \in \Sigma^+$ be a word which is not an inverse Lyndon word and
let $\ICFL(w) = (m_1, \ldots, m_k)$. Let $i$ be an integer
such that $1 < i \leq k$.
Let $r_h, s_h, t_h  \in \Sigma^*$, $a_h, b_h \in \Sigma$ be such that
$m_h = r_ha_hs_h$,
$m_i = r_hb_ht_h$, $a_h < b_h$, $1 \leq h \leq i-1$.
Then, for each $h$, with $1 \leq h < i-1$,
we have $r_{i-1}b_{i-1} \not \in \Fact(m_hm_{h+1})$.
\end{lemma}
\begdim
Let $w \in \Sigma^+$ be a word which is not an inverse Lyndon word and
let $\ICFL(w) = (m_1, \ldots, m_k)$. Let $i$ be an integer
such that $1 < i \leq k$.
Let $r_h, s_h, t_h  \in \Sigma^*$, $a_h, b_h \in \Sigma$ be such that
$m_h = r_ha_hs_h$,
$m_i = r_hb_ht_h$, $a_h < b_h$, $1 \leq h \leq i-1$.

Suppose the lemma were false.
Then we could find $h$, with $1 \leq h < i-1$,
such that $r_{i-1}b_{i-1} \in \Fact(m_hm_{h+1})$.
Therefore, there are $u, v \in \Sigma^*$ such that
$u r_{i-1}b_{i-1} v = m_hm_{h+1}$.
The words $u$ and $m_h$ (resp. $v$ and $m_{h+1}$)
are comparable for the prefix (resp. suffix) order.
Moreover, by Lemma \ref{Lemma1}, $m_h$ (resp. $m_{h+1}$)
is not a prefix (resp. suffix) of $u$ (resp. $v$).
Consequently there are
$r, s \in \Sigma^+$ such that
\begin{eqnarray} \label{eq12}
&& m_h = ur, \quad r_{i-1}b_{i-1} = rs, \quad m_{h+1} = sv
\end{eqnarray}
In addition, $u \not = 1$, otherwise $m_h \in \Pref(m_i)$, in contradiction
with $m_h \ll m_i$.
Therefore, $r$ is a proper nonempty suffix of $m_h$.
Moreover, notice that $r$ is a prefix of $m_i$.
We claim that $r \not \in \Pref(m_h)$.
Indeed, if $r$ were a prefix of $m_h$, it would be a nonempty border
of $m_h$. Thus, on one hand $r \not \in \Pref(m_{h+1})$
by Proposition \ref{bordo4}. On the other hand, $r$ would
be a prefix both of $m_h$ and $m_i$, hence
$r \in \Pref(m_{h+1})$ by Corollary \ref{corollariopreliminary-zero},
a contradiction.

Since $r$ is a proper nonempty suffix of $m_h$ and
$r \not \in \Pref(m_h)$,
we have $r \ll m_h$ which yields
$m_i \ll m_h$, because $r \in \Pref(m_i)$.
This is impossible since $m_h \ll m_i$.
\enddim

\begin{lemma} \label{Lemma3}
Let $w \in \Sigma^+$ be a word which is not an inverse Lyndon word and
let $\ICFL(w) = (m_1, \ldots, m_k)$. Let $i$ be an integer
such that $1 < i \leq k$.
Let $r_h, s_h, t_h  \in \Sigma^*$, $a_h, b_h \in \Sigma$ be such that
$m_h = r_ha_hs_h$,
$m_i = r_hb_ht_h$, $a_h < b_h$, $1 \leq h \leq i-1$.
Then, we have
$r_{i-1}b_{i-1} \not \in \Fact(m_1 \cdots m_{i-1})$.
\end{lemma}
\begdim
Let $w \in \Sigma^+$ be a word which is not an inverse Lyndon word and
let $\ICFL(w) = (m_1, \ldots, m_k)$. Let $i$ be an integer
such that $1 < i \leq k$.
Let $r_h, s_h, t_h  \in \Sigma^*$, $a_h, b_h \in \Sigma$ be such that
$m_h = r_ha_hs_h$,
$m_i = r_hb_ht_h$, $a_h < b_h$, $1 \leq h \leq i-1$.

By contradiction, suppose that
$r_{i-1}b_{i-1} \in \Fact(m_1 \cdots m_{i-1})$.
Thus there are $z, z' \in\Sigma^*$
such that $z r_{i-1}b_{i-1} z' = m_1 \cdots m_{i-1}$.
By Lemma \ref{Lemma2},
for each $h$, with $1 \leq h < i-1$,
we have $r_{i-1}b_{i-1} \not \in \Fact(m_hm_{h+1})$.
Therefore, there is $h$, $1 \leq h \leq i-1$,
such that $m_h \in \Fact(r_{i-1}b_{i-1})$.

Take $h$ minimal with respect to this condition.
Then, there would be $u, v \in\Sigma^*$ such that
$u m_h v = r_{i-1}b_{i-1}$ which implies
$zu m_h vz' = m_1 \cdots m_{i-1}$.
We have $u \not = 1$,
otherwise $m_h \in \Pref(m_i)$, in contradiction
with $m_h \ll m_i$. Thus $h > 1$.
The words $m_{h-1}$ and $u$ are comparable
for the suffix order.
In addition, $m_{h-1}$
is not a suffix
of $u$ by the minimality of $h$.
Hence $u$ would be a nonempty
proper suffix of $m_{h-1}$.
Moreover, $h < i-1$, since $m_{i-1}$ starts with
$r_{i-1}a_{i-1}$.
Notice that $u$ is a prefix of $m_i$.

We now use the same argument as in Lemma \ref{Lemma2}.
We claim that $u \not \in \Pref(m_{h-1})$.
Indeed, if $u$ were a prefix of $m_{h-1}$, then $u$ would be a nonempty border
of $m_{h-1}$. Thus, on one hand $u \not \in \Pref(m_{h})$
by Proposition \ref{bordo4}. On the other hand, $u$ would
be a prefix both of $m_{h-1}$ and $m_i$, hence
$u \in \Pref(m_{h})$ by Corollary \ref{corollariopreliminary-zero},
a contradiction.

Since $u$ is a proper nonempty suffix of $m_{h-1}$ and
$u \not \in \Pref(m_{h-1})$,
we have $u \ll m_{h-1}$ which yields
$m_i \ll m_{h-1}$, because $u \in \Pref(m_i)$.
This is impossible since $m_{h-1} \ll m_i$.
\enddim

\subsection{The Main Result}

Let $w \in \Sigma^+$ be a word which is not an inverse Lyndon word.
Let $\ICFL(w)=(m_1, \ldots, m_k)$. For any suffix $x$ of $m_i$,
$1 \leq i \leq k$, we set $x_w = xm_{i+1} \cdots m_{k}$.
In this section we compare a pair of suffixes $x,y$ of
factors in $\ICFL(w)$ and the corresponding pair of suffixes
$x_w, y_w$ of $w$, with respect to $\lcp$.
First we handle suffixes of the same factor $m_{i}$
(Lemma \ref{suffix-relationship-basic}), then we focus on
suffixes of two different factors $m_{i}, m_{j}$
(Lemma \ref{suffix-relationship-basic2}).

\begin{lemma} \label{suffix-relationship-basic}
Let $w \in \Sigma^+$ be a word which is not an inverse Lyndon word.
Let $\ICFL(w)=(m_1, \ldots, m_k)$.
Let $r, s, t \in \Sigma^*$,
$a, b \in \Sigma$ be such that $m_{i -1} = ras$, $m_i = rbt$, $a < b$,
$1 < i \leq k$.
If $x,y$ are different nonempty suffixes of $m_{i-1}$, then $\lcp(x_w, y_w)= \lcp (xr, yr)$.
\end{lemma}
\begdim
Let $w \in \Sigma^+$ be a word which is not an inverse Lyndon word.
Let $\ICFL(w)=(m_1, \ldots, m_k)$.
Let $r, s, t \in \Sigma^*$,
$a, b \in \Sigma$ be such that $m_{i -1} = ras$, $m_i = rbt$, $a < b$,
$1 < i \leq k$.
Let $x,y$ be different nonempty suffixes of $m_{i-1}$. Set $z = \lcp(x_w, y_w)$.
If $|z| \leq \min\{|xr|, |yr| \}$, then clearly
$\lcp(x_w, y_w)= \lcp (xr, yr)$.

Assume $|z| > \min\{|xr|, |yr| \}$. Thus, the words
$xr, yr$ are comparable for the prefix order.
Let $u \in \Sigma^+$ be such that $yr = xru$ (a symmetric argument applies
if $yr$ is a proper prefix of $xr$). Thus
$|x| < |y|$.
Since $z$ is a prefix of $xrbtm_{i+1} \cdots m_k$
and $|z| > \min\{|xr|, |yr| \} = |xr|$,
there is $v \in \Sigma^*$ such that
$z = xrbv$. Therefore the words $xrb$ and $y$ are comparable for the prefix order,
because they are both prefixes of the same word $y_w$. Hence there is
$v_1 \in \Sigma^*$ such that one of the following two cases holds.
\begin{eqnarray}
y & = & xrbv_1 \label{eq4} \\
xrb & = & yv_1, \quad v_1 \not = 1 \label{eq5}
\end{eqnarray}
Both cases lead to a contradiction.
If Eq. (\ref{eq4}) holds, then $rbv_1$ is a suffix of $m_{i -1} = ras$
and $m_{i -1} \ll rbv_1$, which is impossible.
If Eq. (\ref{eq5}) holds, since $|x| < |y|$ and $v_1 \not = 1$,
we have $y = xr'$, where $r'$ is a nonempty
prefix of $r$. Thus $r'$ is a nonempty border of
$m_{i -1}$ and $r'$ is a prefix of $m_i$, in contradiction
with Proposition \ref{bordo4}.
\enddim

\begin{lemma} \label{suffix-relationship-basic2casodiff}
Let $w \in \Sigma^+$ be a word which is not an inverse Lyndon word and
let $\ICFL(w) = (m_1, \ldots, m_k)$. Let $i, j$ be integers
such that $1 \leq i < j \leq k$.
If $x$ is a nonempty suffix of $m_{i-1}$ and $y$
is a nonempty suffix of $m_{j-1}$ such that $x$ is a proper prefix of $y$, then
$\lcp(x_w, y_w)$ is a prefix of $ym_j$.
\end{lemma}
\begdim
Let $w \in \Sigma^+$ be a word which is not an inverse Lyndon word and
let $\ICFL(w) = (m_1, \ldots, m_k)$. Let $i, j$ be integers
such that $1 \leq i < j \leq k$.
Let $x$ be a nonempty suffix of $m_{i-1}$ and let $y$
be a nonempty suffix of $m_{j-1}$ such that $x$ is a proper prefix of $y$.
Let $r_{j-1}, s_{j-1}, t_{j-1} \in \Sigma^*$,
$a_{j-1}, b_{j-1} \in \Sigma$ be such that $m_{j-1} = r_{j-1}a_{j-1}s_{j-1}$,
$m_j = r_{j-1}b_{j-1}t_{j-1}$, $a_{j-1} < b_{j-1}$.

Set $z = \lcp(x_w, y_w)$.
Since $z$ and $ym_j$ are prefixes of the same word $y_w$, they
are comparable for the prefix order.
By contradiction, assume that $z$ is not a prefix of $ym_j$.
Thus $ym_j$ is a proper prefix of $z$, hence of $x_w$.
Since $ym_j$ and $xm_i \cdots m_{j-1}m_{j}$ are both prefixes of
the same word $x_w$, they are comparable for the prefix order.
Moreover $|xm_i \cdots m_{j-1}m_{j}| > |ym_j|$
because $y$ is a suffix of $m_{j-1}$ and $x$ is nonempty.
Hence there exists $v_j \in \Sigma^+$ such that
\begin{eqnarray} \label{eq10}
ym_jv_j &=& y r_{j-1}b_{j-1}t_{j-1} v_j = xm_i \cdots m_{j-1}m_{j}
\end{eqnarray}
Since $x$ is a proper prefix of $y$, there is $x' \in \Sigma^+$
such that $y = xx'$. Therefore, by Eq. (\ref{eq10}) we have
\begin{eqnarray} \label{eq11}
x' r_{j-1}b_{j-1}t_{j-1} v_j & = & m_i \cdots m_{j-1}m_{j}
\end{eqnarray}
If $|m_j| \leq |t_{j-1} v_j|$, then by Eq. (\ref{eq11})
we have $r_{j-1}b_{j-1} \in \Fact(m_i \cdots m_{j-1})$,
in contradiction with Lemma \ref{Lemma3}.
Hence $|b_{j-1}t_{j-1} v_j| \leq |m_j| < |r_{j-1}b_{j-1}t_{j-1} v_j|$.
Thus, by Eq. (\ref{eq10}),
there are $r'_{j-1} \in \Sigma^+$,
$r''_{j-1} \in \Sigma^*$ such that
$r_{j-1} = r'_{j-1} r''_{j-1}$
and
\begin{eqnarray} \label{eq13}
yr'_{j-1} & =& xm_i \cdots m_{j-1}
\end{eqnarray}
The word $r'_{j-1}$
is a proper prefix of $m_{j-1}$, thus,
by Eq. (\ref{eq13}), $r'_{j-1}$ is a nonempty border of $m_{j-1}$.
Since $r'_{j-1}$ is a prefix of $m_{j}$, this is in contradiction
with Proposition \ref{bordo4}.
\enddim

\begin{lemma} \label{suffix-relationship-basic2}
Let $w \in \Sigma^+$ be a word which is not an inverse Lyndon word and
let $\ICFL(w) = (m_1, \ldots, m_k)$. Let $i, j$ be integers
such that $1 < i < j \leq k$.
If $x$ is a nonempty suffix of $m_{i-1}$ and $y$
is a nonempty suffix of $m_{j-1}$, then
$\lcp(x_w, y_w)$ is a prefix of $ym_j$.
\end{lemma}
\begdim
Let $w \in \Sigma^+$ be a word which is not an inverse Lyndon word.
Let $\ICFL(w)=(m_1, \ldots, m_k)$.
Let $r_{i}, s_{i}, t_{i} \in \Sigma^*$,
$a_{i}, b_{i} \in \Sigma$ be such that $m_{i} = r_ia_is_i$, $m_j = r_ib_it_i$, $a_i < b_i$,
$1 < i < j \leq k$.
Let $r_{i-1}, s_{i-1}, t_{i-1} \in \Sigma^*$,
$a_{i-1}, b_{i-1} \in \Sigma$ be such that $m_{i-1} = r_{i-1}a_{i-1}s_{i-1}$,
$m_j = r_{i-1}b_{i-1}t_{i-1}$, $a_{i-1} < b_{i-1}$.
By Lemma \ref{preliminary-zero}, $r_{i-1}$ is a prefix of $r_i$.

Let $x$ be a nonempty suffix of $m_{i-1}$ and let $y$
be a nonempty suffix of $m_{j-1}$. If $x$ is a proper prefix of $y$, then
by Lemma \ref{suffix-relationship-basic2casodiff} we are done.
Thus assume that $x$ is not a prefix of $y$.
Set $z = \lcp(x_w, y_w)$.
If $|z| \leq |yr_{i-1}|$, then $z$ is a prefix of $ym_j \cdots m_k$ shorter than
$yr_{i-1}b_{i-1}t_{i-1} = ym_j$, thus $z$
is a prefix of $ym_j$.
Assume
$|z| > |yr_{i-1}|$.

Since $z$ is a prefix of $ym_j \cdots m_k = yr_{i-1}b_{i-1}t_{i-1}m_{j+1} \cdots m_k$
and $|z| > |yr_{i-1}|$,
there is $v \in \Sigma^*$ such that
$z = yr_{i-1}b_{i-1}v$. Therefore the words $yr_{i-1}b_{i-1}$ and $x$ are comparable for the prefix order,
because they are both prefixes of the same word $x_w$. Hence there is
$v_1 \in \Sigma^*$ such that one of the following two cases holds.
\begin{eqnarray}
x & = & yr_{i-1}b_{i-1}v_1 \label{eq6} \\
yr_{i-1}b_{i-1} & = & xv_1, \quad v_1 \not = 1 \label{eq7}
\end{eqnarray}
Both cases lead to a contradiction.
If Eq. (\ref{eq6}) holds, then $r_{i-1}b_{i-1}v_1$ is a suffix of $m_{i -1} = r_{i-1}a_{i-1}s_{i-1}$
and $m_{i -1} \ll r_{i-1}b_{i-1}v_1$, which is impossible.
Assume that Eq. (\ref{eq7}) holds. The words $x$ and $y$
are comparable for the prefix order and
$x$ is not a prefix of $y$. Therefore
we have $x = yr'_{i-1}$, where $r'_{i-1}$ is a nonempty
prefix of $r_{i-1}$. Thus $r'_{i-1}$ is a nonempty proper prefix of both
$m_{i -1}$ and $m_i$. Since $x = yr'_{i-1}$, the word $r'_{i-1}$ is a nonempty border of
$m_{i -1}$ and $r'_{i-1}$ is a prefix of $m_i$, in contradiction
with Proposition \ref{bordo4}.
\enddim

Let $w \in \Sigma^+$ be a word which is not an inverse Lyndon word
and let $\ICFL(w) = (m_1, \ldots, m_k)$.
We set
$$\M = \max\{ |m_im_{i+1}| ~|~ 1 \leq i < k\}$$

\begin{proposition} \label{lcpdiversi}
Let $w \in \Sigma^+$ be a word which is not an inverse Lyndon word
and let $\ICFL(w) = (m_1, \ldots, m_k)$.
Let $i, j$ be integers
such that $1 < i < j \leq k$.
If $x$ is a nonempty suffix of $m_{i-1}$ and $y$
is a nonempty suffix of $m_{j-1}$, then
$$\LCP(x_w,y_w) = |\lcp(x_w, y_w)| \leq \M$$
\end{proposition}
\begdim
Let $w \in \Sigma^+$ be a word which is not an inverse Lyndon word
and let $\ICFL(w) = (m_1, \ldots, m_k)$.
Let $i, j$ be integers
such that $1 \leq i < j \leq k$.
Let $x$ be a nonempty suffix of $m_{i-1}$ and let $y$
be a nonempty suffix of $m_{j-1}$.
By Lemma \ref{suffix-relationship-basic2},
$\lcp(x_w, y_w)$ is a prefix of $ym_j$, hence
$$\LCP(x_w,y_w) = |\lcp(x_w, y_w)| \leq |ym_j| \leq |m_{j-1}m_j| \leq \M$$
\enddim

\begin{proposition} \label{lcpfattori}
Let $w \in \Sigma^+$ be a word which is not an inverse Lyndon word
and let $\ICFL(w) = (m_1, \ldots, m_k)$.
For each nonempty proper factors $u, v$ of $w$, we have
$$\LCP(u,v) = |\lcp(u,v)| \leq \M$$
\end{proposition}
\begdim
Let $w \in \Sigma^+$ be a word which is not an inverse Lyndon word
and let $\ICFL(w) = (m_1, \ldots, m_k)$.
Let $u, v$ be nonempty proper factors of $w$.
Let $u_1, u_2, v_1,v_2 \in \Sigma^*$ be such that
$w = u_1uu_2 = v_1vv_2$.
Let $x$ be a nonempty suffix of $m_{i-1}$ and let $y$
be a nonempty suffix of $m_{j-1}$
such that $uu_2 = x_w$, $vv_2 = y_w$, with $1 < i \leq k$,
$1 < j \leq k$.
If $i = j$, then by Lemma \ref{suffix-relationship-basic}
we have
$$\LCP(u,v) = |\lcp(u,v)| \leq |\lcp(x_w, y_w)| \leq |m_{i-1}m_i| \leq \M$$
If $i \not = j$, then by Proposition \ref{lcpdiversi}, we have
$$\LCP(u,v) = |\lcp(u,v)| \leq |\lcp(x_w, y_w)| \leq \M$$
\enddim


\end{document}